\documentstyle[11pt,aaspp4]{article}

\lefthead{F. La Franca and S. Cristiani}
\righthead{The QSO evolution from the {\it HBQS}}

\begin{document}

\title{ERRATUM: ``The QSO evolution derived from the {\it HBQS} \\ 
and other complete QSO surveys 
\footnote{Based on data of the ESO Key-Programme ``A Homogeneous Bright
QSO Survey''}''[ASTRON. J., 113, 1517 (1997)]}

\author{Fabio La Franca}
\affil{Dipartimento di Fisica, Universit\`a degli studi ``Roma Tre''\\
Via della
Vasca Navale 84, I-00146 Roma, Italy\\
{\rm Electronic mail: lafranca@amaldi.fis.uniroma3.it}}

\and
\author{Stefano Cristiani}
\affil{Dipartimento di Astronomia, Universit\`a degli studi di Padova \\
Vicolo dell'Osservatorio 5, I-35122 Padova, Italy\\
{\rm Electronic mail: cristiani@astrpd.pd.astro.it}
\\~~~~\\~~~~~\\~~~~~\\
}

The equation used for the simulations of the LDLE models is incorrectly
reported. The correct equation for the LDLE evolution of the break magnitude
is $ M_B(z) = M_B^{\ast}(2) - 2.5k\log[(1+z)/3]$ where $M_B^{\ast}(2)$ is the
break magnitude at $z=2$ and
\begin{eqnarray}
\nonumber for~M_B \leq M_B(z)&:&~ k = k_1 + k_2 (M_B-M_B(z))e^{-z/{.40}} \\
\nonumber for~M_B > M_B(z)&:&~ k = k_1 . 
\end{eqnarray}
In Table 2 the values of $M^{\ast}_B$ for the LDLE models C and M should be
substituted by the values for $M_B^{\ast}(2)$. They should read, for model C,
-26.3; for model M, -26.9. This change does not concern all the other models
reported in Table 2 which use the standard PLE parameterization reported in
the paper. We thank L. Wisotzki for turning our attention toward this point.

\end{document}